\documentclass[]{interact}
\articletype{Preprint Version}

\usepackage{epstopdf}
\usepackage[caption=false]{subfig}
\usepackage{multirow}
\usepackage{graphicx}
\usepackage{booktabs}
\usepackage{makecell}
\usepackage{comment}
\usepackage{enumitem}

\usepackage[natbibapa,nodoi]{apacite}
\theoremstyle{plain}

\theoremstyle{definition}

\theoremstyle{remark}

\begin{document}


\title{ChatGPT on the Road: Leveraging Large Language Model-Powered In-vehicle Conversational Agents for Safer and More Enjoyable Driving Experience}

\author{
\name{Yeana Lee Bond\textsuperscript{a†}, Mungyeong Choe\textsuperscript{b†}\thanks{†These authors have contributed equally to this work and share the first authorship.}, Baker Kasim Hasan\textsuperscript{a}, Arsh Siddiqui\textsuperscript{a}, Myounghoon Jeon\textsuperscript{ab$^\ast$}\thanks{*Corresponding author: Dr. Myounghoon Jeon. Email: myounghoonjeon@vt.edu}}
\affil{\textsuperscript{a}Computer Science, Virginia Tech, Blacksburg, Virginia, USA;} \textsuperscript{b}Industrial and Systems Engineering, Virginia Tech, Blacksburg, Virginia, USA
}

\maketitle 

\begin{abstract}
Studies on in-vehicle conversational agents have traditionally relied on pre-scripted prompts or limited voice commands, constraining natural driver-agent interaction. To resolve this issue, the present study explored the potential of a ChatGPT-based in-vehicle agent capable of carrying continuous, multi-turn dialogues. Forty drivers participated in our experiment using a motion-based driving simulator, comparing three conditions (No agent, Pre-scripted agent, and ChatGPT-based agent) as a within-subjects variable. Results showed that the ChatGPT-based agent condition led to more stable driving performance across multiple metrics. Participants demonstrated lower variability in longitudinal acceleration, lateral acceleration, and lane deviation compared to the other two conditions. In subjective evaluations, the ChatGPT-based agent also received significantly higher ratings in competence, animacy, affective trust, and preference compared to the Pre-scripted agent. Our thematic analysis of driver-agent conversations revealed diverse interaction patterns in topics, including driving assistance/questions, entertainment requests, and anthropomorphic interactions. Our results highlight the potential of LLM-powered in-vehicle conversational agents to enhance driving safety and user experience through natural, context-rich interactions.
\end{abstract}


\section{Introduction}

\begin{figure*}[t]
\label{tab:mostcommonactivity}
\centering
\begin{tabular}{c|c|c}
\cline{1-3}
\includegraphics[width=0.26\textwidth,height=2.7cm]{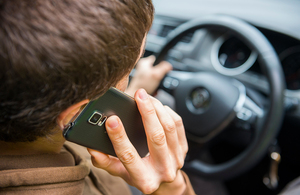} & 
\includegraphics[width=0.28\textwidth,height=2.73cm]{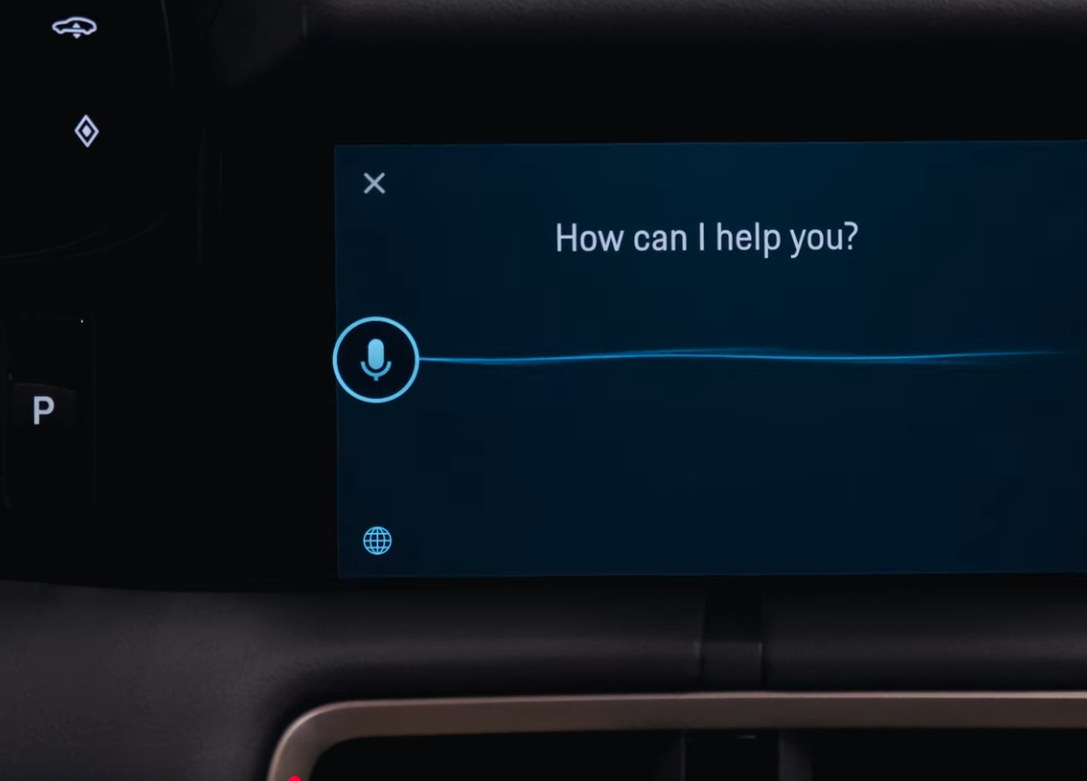} & 
\includegraphics[width=0.35\textwidth,height=2.7cm]{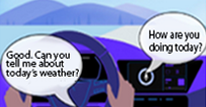} \\
\cline{1-3}
\multicolumn{1}{c|}{(a)} & \multicolumn{1}{c|}{(b)} & \multicolumn{1}{c}{(c)} \\
\cline{1-3}
\end{tabular}
\caption{(a) A driver conversing with someone via a mobile phone held by the driver's right hand while using the other hand for driving simultaniously (b) An example of a Voice User Interface capable of processing human speech and verbally interact with a user based on pre-defined, canned responses often with a corresonponding visual display(c) A visualized example of an LLM-powered in-vehicle conversational agent, CARA in our present study, capable of taking multi-turn conversations in human natural language and receiving a question about today's weather, ``Can  you tell me about today's weather?" from a driver whose both hands are on the wheel}

\end{figure*}

The rapid evolution of Artificial Intelligence (AI), propelled most recently by the advent of large-scale, pre-trained language models, has fundamentally reshaped human-machine interaction \citep{pal2023future}. This transformation is aligned with a long-standing aspiration in Natural Language Processing (NLP), Human-Computer Interaction (HCI), and Intelligent User Interface (IUI) research \citep{schmidt2021introduction,rege2024talking,volkel2020intelligent} to offer Adaptive User Interfaces (AUIs) \citep{jeon2014effects,rittger2022adaptive} to enhance performance, affectivity, and naturalness by leveraging a user’s human language in its natural form \citep{bernsen2001exploring,geutner2002design,jalil2022introduction,liao2023,maybury1998readings,mctear2002spoken,schmidt2021introduction}. Rather than rigid, command-based instructions \citep{miller2018voice}, researchers in spoken dialogue systems and Conversational User Interfaces (CUIs) \citep{geutner2002design,hofmann2014comparison,large2019,weng2016conversational,zue2000conversational} developed related technologies under various names---including chatbots \citep{gao2018neural,klopfenstein2017rise,landay2019conversational}, virtual assistants \citep{braun2019your,braun2019improving,lai2000conversational,eyben2010emotion}, and voice agents \citep{dong2020female,ji2022designing,Lee2019,park2024effects,seaborn2021voice}---even before the surge of the OpenAI’s models. The surge began with the release of GPT-3.5 in November 2022 \citep{openai_gpt35base}, followed by GPT-3.5-Turbo in March 2023 along with Whisper \citep{openai_gpt35turbo}, and soon after GPT-4 in the same month \citep{openai_gpt4}.

From command-based voice assistants such as Microsoft’s Cortana or Apple’s Siri \citep{hoy2018alexa} to LLM-powered Conversational AI such as CareCall deployed across 20 municipalities in South Korea powered by Naver’s hyperCLOVA \citep{clovaCareCall2022Seoul, jo2023understanding}, Duolingo’s Duolingo Max \citep{duoLingoMax2023}, Khan Academy’s Khanmigo \citep{khanPhi2024}, and Amazon’s Rufus \citep{amazon2024rufus}, a wider range of uses for LLM-powered Conversational AI across numerous industries have been found, including in healthcare, education, and customer service. In the automotive industry, several companies announced the development of virtual voice assistants in their vehicles. Since then, BMW collaborated with Amazon’s Alexa \citep{bmw2024Amazon, alexaBMW} while General Motors partnered with Microsoft’s Azure Cloud service \citep{weatherbed2023chatgpt}. Mercedes-Benz also tested ChatGPT to enhance the underlying features of user experience via their voice assistant \citep{betaMercedes2024chatGPT}.

\begin{figure*}[t]
    \centering
    \includegraphics[width=\textwidth]{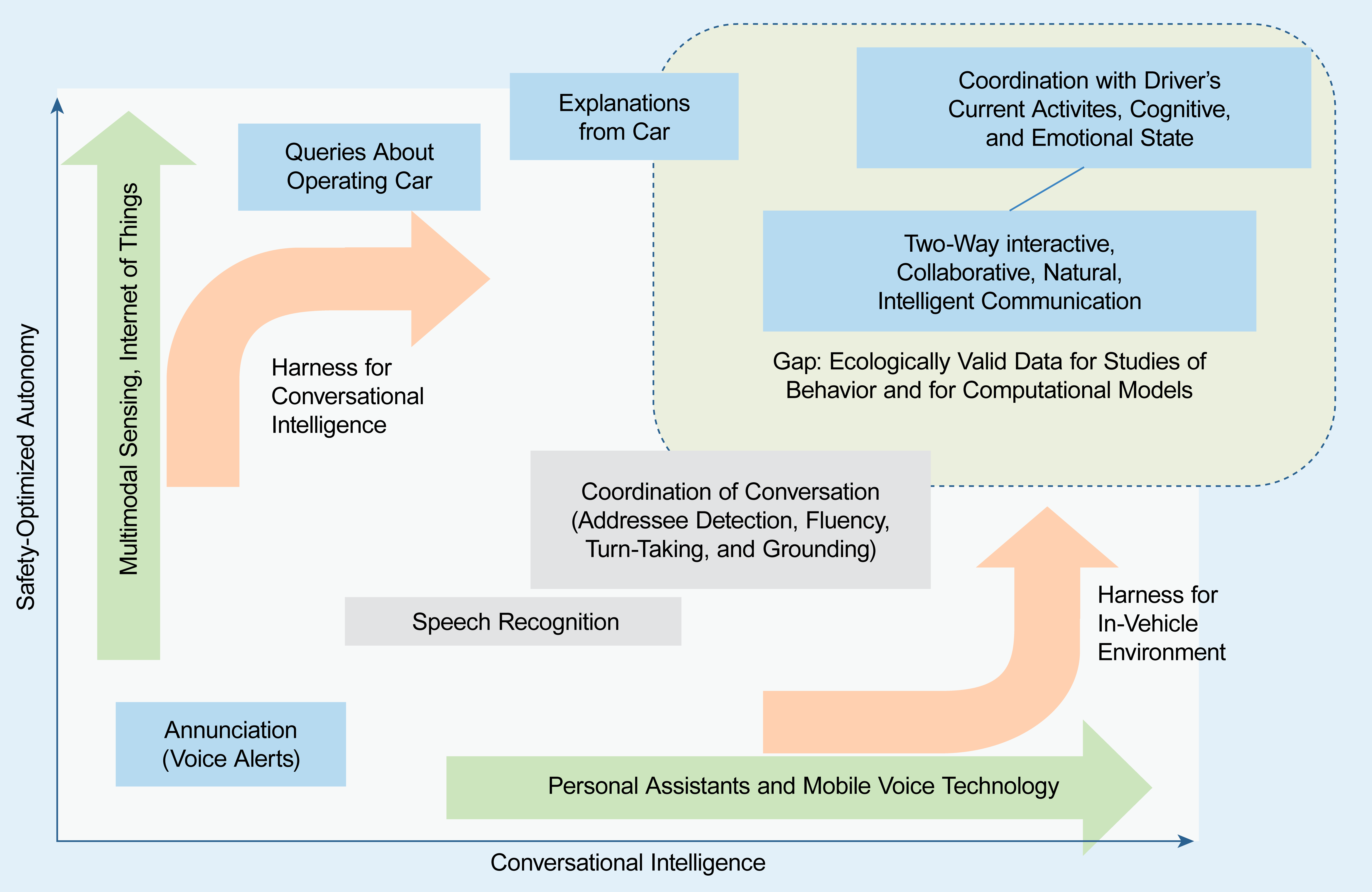}
    \caption{Future directions of in-vehicle conversational and intelligent dialogue systems \citep{weng2016conversational}}
    \label{futureCID}
\end{figure*}

When designing an in-vehicle intelligent dialogue system, researchers have identified three critical factors specific to the driving context: (1) the driver, (2) the environment, and (3) the automotive industry \citep{weng2016conversational}. Our present study focused on human users driving with the automation Level 0, which means no driving automation according to the Society of Automotive Engineers (SAE) Levels of Driving Automation \citep{saej3016}. With respect to the first factor (driver), we included emotions as part of the driver state, as emotional states significantly influence driving behavior. Although driving is often considered as a single task \citep{hsieh2010effect}, it is inherently complex, and driver’s emotional state can impact safety and lead to consequences \citep{jeon2017emotions, steinhauser2018effects}. 

Users often experience emotional responses to system designs, regardless of their stated preferences during the design process \citep{andre1995users}. In terms of the second and third factors (the environment and the automotive industry respectively), we assumed that road events after which a corresponding, pre-scripted response was played by the ChatGPT application might also potentially influence driver emotions according to supporting studies and research that considered emotional factors or conversational interactions in vehicles \citep{braun2019improving,braun2021affective,choe2023see,choi2018designing,eyben2010emotion,Lee2019,lee2022systematic,jeon2014effects,jeon2024novel,kanevsky2008telematics,sheller2004automotive,strassmann2023,steinhauser2018effects,wang2021vehicle,wang2022conversational,zepf2019towards,zepf2020driver}. We recognized that natural emotional engagement would likely occur not only in response to the in-vehicle conversational agent’s empathic utterances but also throughout general conversational interaction, as speech-based systems are known to evoke and convey a wide range of emotions \citep{weng2016conversational}. In-vehicle emotions, therefore, emerge from the dynamic interplay among vehicle, road, and driver \citep{sheller2004automotive,weng2016conversational,zepf2019towards}. 

Researchers in Affective Computing have shown that drivers’ emotional states, whether negative or positive, are core components of the overall driver state and can influence driving performance \citep{jeon2014effects, sheller2004automotive, steinhauser2018effects, zepf2019towards, zepf2020driver}. However, no consistent relationship has been identified between specific emotions (e.g., frustration or sadness versus happiness) and improved or diminished driving performance \citep{jeon2014effects}. Consequently, positive emotions do not always correspond to superior driving performance, nor do negative emotions necessarily lead to worse performance \citep{jeon2014effects}. Thus, the present study examines those critical factors \citep{weng2016conversational} and explores how the simulated affective empathy from the in-vehicle conversational agent influences driver behavior, guiding the design of LLM-powered in-vehicle conversational interfaces. For two-way conversational interactions between the ChatGPT-based agent and drivers, we used OpenAI’s ChatGPT application. This approach supports our goal of enhancing the driver experience through Safety-Optimized Autonomy and Conversational Intelligence, as illustrated in Figure \ref{futureCID}.

Some drivers who prefer minimal distraction with in-vehicle agents still favor brief, command-based dialogues to ensure driver safety \citep{hofmann2014comparison} because conversational interaction in human natural language tends to be more time-consuming due to the need for turn-taking \citep{lai2000conversational,schmandt1994voice}. This means that conversational interaction may not always be an efficient way to communicate with machines or computers. However, its effectiveness can be modulated when the interaction is with an intelligent counterpart \citep{drews2008passenger} such as an in-vehicle conversational agent intelligent enough to be aware of the environments inside and outside a vehicle and able to predict a driver's cognitive and emotional states depending on the environmental changes \citep{weng2016conversational}.

\cite{liang2014looming} analyzed observations from a naturalistic field study with 100 vehicles and found that glances away from the road lasting 1.7 seconds or longer significantly increase the risk of an accident. Furthermore, verbal attention and memory can interfere with visual processing required for driving \citep{Recarte2000Effects}. Based on these findings, Voice User Interfaces (VUIs) align well with Multiple Resources Theory (MRT) in that using different modalities can promote time sharing of better multi-tasking \citep{Large2017,wickens2002multiple}. However, users are often unaware of underlying information-processing mechanisms that balance performance and preference \citep{andre1995users}. At times, while driving, they may engage in other activities such as conversation, not for efficiency of driving, but to satisfy personal or affective needs \citep{taylor2006americans}. Indeed, some drivers enjoy conversing while driving before and after the in-car connectivity via Bluetooth was an option \citep{dmeautomotive2014caraoke,taylor2006americans}, perhaps viewing the two activities as one familiar, low-effort, and coupled activity.

Human factors researchers have noted that user preferences do not always correspond to actual performance outcomes \citep{andre1995users}. In other words, users may prefer certain features in a human-machine interface that do not provide them with the best performance. Preferences may be influenced instead by factors such as novelty, familiarity, or low effort \citep{andre1995users}. These subjective factors may ultimately impede improvements in performance and efficiency \citep{andre1995users}. Therefore, it is important to evaluate not only drivers’ preferences but also their performance outcomes when using this new technology \citep{andre1995users}. Taking these as motivating guidelines, we evaluated driving performance and subjective ratings of perceptions, trust, and overall driving experience as well as emotion. We also recorded utterances during driver interaction with the in-vehicle conversational agent in a driving context with the automation Level 0 \citep{saej3016}, which remains the most common level worldwide. Using the novel approach combining the most common level of automation worldwide (the environment outside a vehicle) with in-vehicle conversational agents powered by LLMs, we compared the ChatGPT-based agent with a no-agent baseline and a pre-scripted agent (driver-agent interactions made inside a vehicle). 

By answering the following research questions, we aim to share insights that will help researchers, stakeholders, designers, and the automotive industry create user-centered and safe driving environments in our present work.

\begin{itemize}[leftmargin=15pt, topsep=12pt, partopsep=0pt, parsep=0pt]
    \item \textbf{RQ1}. How do different types of in-vehicle agents \textit{(ChatGPT-based, Pre-scripted, or No Agent)} influence driving performance?
    \item \textbf{RQ2}. How does the ChatGPT-based conversational agent influence drivers’ ratings on subjective experiences compared to the pre-scripted agent?
    \item \textbf{RQ3}. How do drivers’ positive and negative affect scores vary across the two different agent types after experiencing them?
    \item \textbf{RQ4}. What conversation topics or patterns emerge during drivers' interactions with the ChatGPT-based agent?
\end{itemize}

\section{Related Work}

\subsection{Efforts Toward Intelligent In-Vehicle Agents/Systems}
In 1996, Mercedes-Benz introduced Linguatronic, the first voice-controlled hands-free system for in-car mobile phones, \citep{heisterkamp2001linguatronic}. Despite widespread announcements by car manufacturers, speech recognition remained a significant challenge for researchers and developers, particularly in implementing voice-operated Command \& Control systems \citep{heisterkamp2001linguatronic}, not to mention the even greater difficulty in achieving error-free, multi-turn conversational interactions in human natural language \citep{kanevsky2008telematics} in the early 2000s. 

Unlike the Artificial Passenger paradigm introduced in 2001 by an IBM researcher \citep{kanevsky2008telematics}, \cite{geutner2002design} introduced the concept of ``a human driver driving together with an intelligent conversational agent" via the Virtual Intelligent Co-Driver (VICO) project in 2002. As part of a European research initiative, VICO was a prototype designed to enable intelligent conversational interaction between human users and digital systems \citep{bernsen2001exploring,geutner2002design}. In the following year, a multimodal conversational interface prototype was implemented in the Ford Model U Concept Vehicle, shown at the 2003 North American International Auto Show in Detroit \citep{meixner2017automotive}. 

As in-vehicle connectivity advanced, and the integration of nomadic devices and internet-based communication became more feasible, companies such as Ford began to push smartphone integration with voice recognition capabilities as a core part of the in-car experience \citep{meixner2017automotive}. To this day, even with emerging resources such as LLMs, most in-vehicle systems still operate under the traditional Command \& Control paradigm. Such systems typically require the memorization of commands mapped to specific functions \citep{miller2018voice}, which are activated when the user expresses a voice command in a certain wording. To address limitations of rudimentary command and control-based interfaces, automotive companies and suppliers have been pursuing research and development of the next generation of in-vehicle speech dialogue systems \citep{meixner2017automotive}. 

Intelligent User Interfaces (IUIs) aim to improve the efficiency, effectiveness, and naturalness of human-machine interaction by representing, reasoning, and acting on models of the users, domain, task, discourse, and multimodal input such as human natural language and gesture \citep{maybury1998readings, volkel2020intelligent}. The  ``intelligent agent" has evolved to encompass the ability to engage in conversation in human natural language \citep{bernsen2001exploring,kanevsky2008telematics}. From an HCI perspective, intelligent agents have traditionally been conceptualized in four roles: (1) as data-entry components of a system, (2) as dialogue partners, (3) as a medium for communication between users and computers, and (4) as tools or instruments to accomplish specific tasks \citep{maybury1998readings}. Our study aligns closely with the second perspective---positioning the agent as a dialogue partner through the application of GPT-4 \citep{openai_gpt4}.

\subsection{Agents in UIs and as Applications}

We have several key adjectives that modify the word, ``agents": The three types are ``In-vehicle", ``Conversational," and ``LLM-powered" agents. An ``agent" is an extension of a human user \citep{kay1990user}, a computing system, and software as a product \citep{lee2022systematic}, and the agent communicates for use of an application via a UI. In Section 1, several related UIs such as IUI, AUI, and CUI were mentioned. In the automotive industry, traditionally, a Graphical User Interface (GUI) and a VUI have coexisted as illustrated in the image (b) of Figure \ref{tab:mostcommonactivity}. Yet, we considered the latter would be the umbrella concept of the UI type that closely overlaps with the concept of CUIs \citep{weng2016conversational} in the present study.

Note that the term, ``assistant" has been used in place of “agent” in previous studies. Although the intended meaning remains similar \citep{lee2022systematic}, we see the possibility that the scope of the two terms may not persist the same way they were due to approaches considering paradigms such as Agentic AI, Autonomous AI \citep{hosseini2025role}, and Software-Defined Vehicles \citep{liu2022impact}. 

Nowadays, Generative AI agents are perceived as applications  powered by models and connected with tools that can extend the knowledge and capabilities of agents in various ways \citep{wiesinger2024agents}. In the 2024 Consumer Electronics Show, Volkswagen announced that they would integrate ChatGPT into some of their vehicle models via a partnership with Cerence, an auto software company \citep{cerence2024volks}. Recently, the company presented a Gemini-based virtual assistant hosted on Google Cloud via the myVW app \citep{googlecloud2024volkswagen}. Like this recent example and our approach using mobile devices, drivers can access an assistant or agent inside their vehicles. 

Using a motion-based driving simulator and introducing unexpected road events that could affect driver emotions and safety, our experiment incorporated the three critical factors identified for in-vehicle intelligent dialogue system \citep{weng2016conversational}. Our ChatGPT-based agent was designed to be perceived as a standalone dialogue partner, allowing observations to directly reflect driver–agent interaction. Now that related conceptual frames and boundary conditions of the present study are defined, the rest of the literature survey was focused around the our agent's two main anthropomorphic capabilities, the abilities to have a conversation and to express affective empathic responses in human natural language with multiple back-and-forth turns with a user.

\subsection{Anthropomorphic Capabilities of LLM-Powered Agents in AUIs}

The level of conversational capabilities of intelligent agents could influence user-perceived levels of qualities and abilities such agents were equipped with. At the intersection of Natural User Interfaces \citep{jain2011future} and IUIs \citep{maybury1998readings,rittger2022adaptive,schmidt2021introduction}, researchers can  leverage advances in natural language understanding, dialogue management, and generation to interpret intent, maintain context, and produce human-like responses that are better performing in emotion awareness than human evaluators by employing LLMs \citep{elyoseph2023chatgpt} than before. Definitions of NUIs \citep{jain2011future} and IUIs often overlap significantly with those of AUIs, as they can be seen as variations of NUIs emphasizing different aspects of human-to-human interaction \citep{zue2000conversational}. These systems use data collected \citep{feeney1977adaptive,kanevsky2008telematics,ji2022designing} from user behaviors and aim to learn user behaviors and preferences, often in natural human-to-human communication through user interfaces. 

This subsection reviews the literature on the two adaptive components of LLM-powered agents, which correspond to their anthropomorphic capabilities as framed by AUI research and the Computers as Social Actors (CASA) paradigm \citep{nass1994computers}.

\subsubsection{Bidirectional, Multi-Turn Conversational Interaction}

In 1950, Alan Turing in his seminal paper ``Computing Machinery and Intelligence.” suggested that within 50 years a computer would pass a comparison test determining if it is a human or a machine \citep{turing2021computing}. The very first such talking bot, a conversational agent, ELIZA dates back to 1966 \citep{weizenbaum1966eliza}. ELIZA simulated the interview style of a mid-20th-century psychotherapist by relying solely on scripted pattern-matching rules rather than any representation of world knowledge, yet it remarkably sustained multi-turn dialogue and gave users the uncanny sense of being heard and understood \citep{weizenbaum1966eliza}. This result demonstrates that even minimal conversational cues can lead users to interact with computing systems as if they were human interlocutors \citep{nass1994computers}.

In 2001, IBM developed software called ``Artificial Passenger," a telemetic device designed to make long driving safer and more bearable by talking to it \citep{kanevsky2008telematics}. They did not use the term, ``agent," but Artificial Passenger was specialized to converse with drivers while allowing them to change radio stations and make jokes to prevent them from drowsing off behind the wheel \citep{kanevsky2008telematics}. Like this, one of the reasons why speech technology has begun to receive more attention is that it would be a good fit for pervasive computing solutions, and pervasive computing devices are conventionally used in hands-busy, eye-busy settings such as driving \citep{geutner2002design,kern2009design,lai2000conversational}. 

Voice-based systems have been shown to reduce a driver’s look-away time and lessen spatial and cognitive demands on the driver \citep{kanevsky2008telematics,miller2018voice,Monk2023Visual}. Human speech is so innate to humans as breathing that we talk to our pets or even to plants \citep{lai2000conversational}. In fact, it is also one of the quintessential markers of humanness \citep{Large2017}, which is why voice-based conversational agents can carry anthropomorphism. Furthermore, \cite{atchley2014strategically} showed that engaging in a verbal task improved performance, even when a driver was fatigued following a long drive around 80 to 90 minutes. They acknowledged that all drivers reported feeling less engaged by the end of the drive, but showed that the drivers in the late verbal task group in which drivers engaged in the verbal task only in the last block exhibited improved lane keeping \citep{atchley2014strategically}.

A study that explored an in-vehicle agent's speech style and embodiment reported that the participants preferred the conversational robot in-vehicle agent for reasons including it being `friendly,' `natural,' and having an `emotional' touch \citep{Lee2019}. However, the study was conducted in the context of full automation with which the seated driver does not perform driving \citep{saej3016}. It also found that participants experienced the highest level of social presence with a conversational, embodied in-vehicle robot agent. In contrast, our study did not incorporate any form of embodiment, thereby eliminating potential visual confounding factors.

``Talking" to machines in a human language through natural, multi-turn dialogues has been redefining the way users interact with technology in various contexts. As part of in-vehicle system factors, a voice-based agent's genders \citep{dong2020female}, speech styles \citep{Lee2019,wang2022conversational}, emotional intelligence to drivers \citep{li2022intelligent, braun2019improving, choe2023see}, or user trust \citep{park2024effects} toward voice-based agents were explored to find factors between driver perceptions, preferences, and driving performance. These studies employed rule-based or pre-scripted communication systems lacking turn-taking capability, functioning merely as one-way announcements triggered at fixed points during the driving sequence. In contrast, our study featured an agent capable of generating utterances that enable back-and-forth conversational interactions.

From here, a conversational agent is defined as a human-computer dialogue system that engages users in turn-by-turn exchanges using human natural language, and we assert that such in-vehicle systems can influence the overall driving experience.

\subsubsection{Empathic Responses in Human Natural Language}

As \cite{cacioppo1993psychophysiology} said, “Emotions guide, enrich, and ennoble life; they provide meaning to everyday existence and underlie the value placed on life and property." Emotion is an essential aspect of the human experience, and many researchers consider emotions and empathy expressed in human natural language in interactions between human users and machines such as \cite{braun2019your,braun2019improving,braun2021affective,choe2023emotion,elyoseph2023chatgpt,eyben2010emotion,garg2022last,hsieh2010effect,huang2025influencing,jeon2024novel,kanevsky2008telematics,large2019,meng2024,pamungkas2019emotionally,strassmann2023,wang2021vehicle}. Empathy, a psychological process resulting in congruent feeling with another individual's situations \citep{eisenberg2001origins}, is at the core of human emotional intelligence in terms of fostering personal connection via social bond \citep{garg2022last,wang2022conversational}. 

Expressing empathy in human natural language is one of the anthropomorphic capabilities that LLM-powered agents can offer, but one might wonder why empathic responses were considered in our present study. According to researchers supporting the CASA paradigm, agents must have personalities to be perceived as believable \citep{nass1995can}, and human-computer relationships fundamentally resemble human-human relationships \citep{nass1994computers}. \cite{nass2005improving} were among the first to investigate an assistant capable of adapting to a user's emotional state, and their findings indicated that participants experienced fewer accidents, exibitited improved attention, and showed a greater willingness to communicate when the system’s voice mirrored the driver’s emotions. This is affective empathy, which involves vicarious emotional experiences \citep{paiva2021empathy}. This emotional resonance can occur automatically and is often referred to as ``vicarious affect," where the empathizer shares in the recipient's feelings as if experiencing them firsthand \citep{paiva2021empathy}. 

Expanding the CASA paradigm, \cite{braun2019your} emphasized that to gain broader acceptance, an agent must meet user expectations for social interaction. Empathy is a factor that can influence conversational interactions in VUIs, which extends to CUIs as well, and therefore should be considered in designing affective conversational agents in general. A comparative study with 60 participants using induction of negative emotion states such as anger or sadness concluded that empathic speech was preferred to visual feedback such as ambient light by the users \citep{braun2019improving}. Another study in which Evoked Response Potential and functional Magnetic Resonance Imaging of 20 participants were recorded showed that anger-mirrored speech output contributed to increasing alertness levels of drivers and thus decreasing distracted driving \citep{hsieh2010effect}. These empirical studies done in the context of driving before the ChatGPT era \citep{Aalst2023Welcome} supported the idea that among cognitive and affective empathy styles, the latter should be taken for our present study. 

Lastly, a study conducted by Microsoft researchers on user expectations for affective conversational agents revealed that language, communication, and emotional support were the most preferred aspects of affective empathy, as voted by 84.5\% of the 745 survey respondents \citep{hernandez2023affective}. This preference and its aspects reflect the inherently human qualities of interaction, such as empathy, and underpin both our study and findings from the CASA paradigm, which highlight the importance of studying in-vehicle assistants adapting to a user’s emotional state while driving \citep{nass1995can,jeon2014effects}. These insights reinforce our choice of affective empathy, as we believe that it enhances naturalness---one of the core attributes of anthropomorphism of LLM-powered conversational agents.

\section{Current Study and Hypothesis}
In-vehicle intelligent agents have attracted interest for their ability to enhance user experience \citep{rege2024talking}, reduce driver stress \citep{li2020effects}, and by providing empathic feedback \citep{braun2021affective,braun2019improving,choe2023emotion,nadri2022empathic,zepf2020driver} and delivering critical driving information to improve safety \citep{wang2022conversational}. These recent studies attempted to conceptualize or emulate aspects of in-vehicle conversational agents especially helpful in situations prone to cause negative emotions such as anger or frustration, which can negatively impact driving performance and overall driving experiences \citep{trick2012fleeting, deffenbacher2002driving, jeon2014effects}. For this reason, we included hazardous road events that could evoke such negative emotions in drivers when designing our experiment to directly compare the ChatGPT-based agent with the pre-scripted one and examine differences in perceptions and performance of drivers as well as their trust levels toward each agent type.

\cite{gao2018neural} stated that one of the unified views toward Conversational AI would be dialogue for optimal decision-making. The researchers identified three roles for conversational systems: (1) question-answering agents, (2) task-oriented dialogue agents, and (3) social bots. These roles align with the four functions of intelligent agents conceptualized by \cite{maybury1998readings}, suggesting that LLM-powered conversational agents should encompass all of these roles and hinting that bidirectional conversation is even the key to exploring other roles of Conversational AI. However, most recent research efforts remain largely conceptual, with limited or no direct user evaluations while suggesting ChatGPT-based agents' roles in tuned with the previously identified ones in intelligent conversational systems \citep{gao2018neural,maybury1998readings}. 

\citet{chanmas2024driving} made progress in the integration of NLP by incorporating pre-generated AI messages in a simulator-based driving assessment; however, their system did not support continuous two-way dialogue. Similarly, \cite{huang2025influencing} employed the Qwen2 model locally to examine the impact of AI assistants’ verbal emotions matching driver emotions for driving safety, yet their work also lacked genuine two-way conversational interaction. With the rise of OpenAI’s ChatGPT-3.5, researchers have begun exploring its potential impact on intelligent vehicles, sharing perspectives on their roles in the field \citep{du2023chat}, preliminary applications as a “co-pilot” for autonomous co-driving for trajectory routing \citep{wangshiyi2023}, and anthropomorphized traits experienced in the use of ChatGPT \citep{simas2024human}. 

Despite these efforts in current research, three key gaps remain. First, most studies did not focus on bidirectional  conversational interactions by enabling dynamic, multi-turn dialogue initiated by the driver. Second, few investigations provide direct comparisons between LLM-powered in-vehicle conversational agents and both pre-scripted and no-agent baselines. Third, while some work suggests potential safety benefits or improvements in user experience, there is a lack of rigorous evaluation of how free-form, LLM-powered conversational agents influence driving performance, road safety, and subjective perceptions. Addressing these gaps is essential for understanding and harnessing the potential of Conversatioanl AI. 

Using OpenAI's GPT-4 \citep{openai_gpt4}, we explored the potential of an LLM-powered in-vehicle agent designed to incorporate affective empathy and enable bidiretional, multi-turn conversational interaction between the agent and a driver in the context of non-automated driving, Level 0 \citep{saej3016}. Rather than relying on one-dimensional prompts, our system allows users to converse with the agent while driving experiencing on-road events in virual scenarios, mirroring natural user interactions. We hypothesized that flexibility and adaptiveness of LLM-powered in-vehicle conversational agents may not only enhance user experience but also have measurable effects on driving behavior. Through a within-subject experiment with three agent conditions (\textit{No agent}, \textit{Pre-scripted agent}, and \textit{ChatGPT-based agent}), we assessed whether an LLM-powered in-vehicle conversational agent can provide a safer and more engaging driving experience.

\section{Methods}
\subsection{Participants}
40 drivers (24 male and 16 female participants) were recruited for the present study. The age of the participants ranges from 19 to 49 years, and the average age is 23.95 (\textpm 5.37). This present work has been reviewed and approved by the Institute of Review Board at [Hidden for review] (approval number: [Hidden for review]). 
Upon receiving approval, the experiment was advertised via physical paper flyers posted on campus and in the immediately surrounding community. The electronic flyers were also distributed using email listservs. Participants were compensated for their participation at a rate of \$10/h or 0.5 credit per hour for one course they chose. The inclusion criteria are: (1) 18 years or older, (2) have at least two years of driving experience with a valid driver's license, (3) a native or fluent English speaker, (4) with normal or corrected normal vision/hearing, (5) with motor skills with hands and feet to drive, (6) willing to wear a lavalier microphone to record verbal utterances and to agree to recording of their facial imagery via a webcam, and (7) no chance of pregancy or wearing an implant such as a pacemaker.

\subsection{Apparatus}

\begin{figure*}[t]
    \centering
    \includegraphics[width=\textwidth]{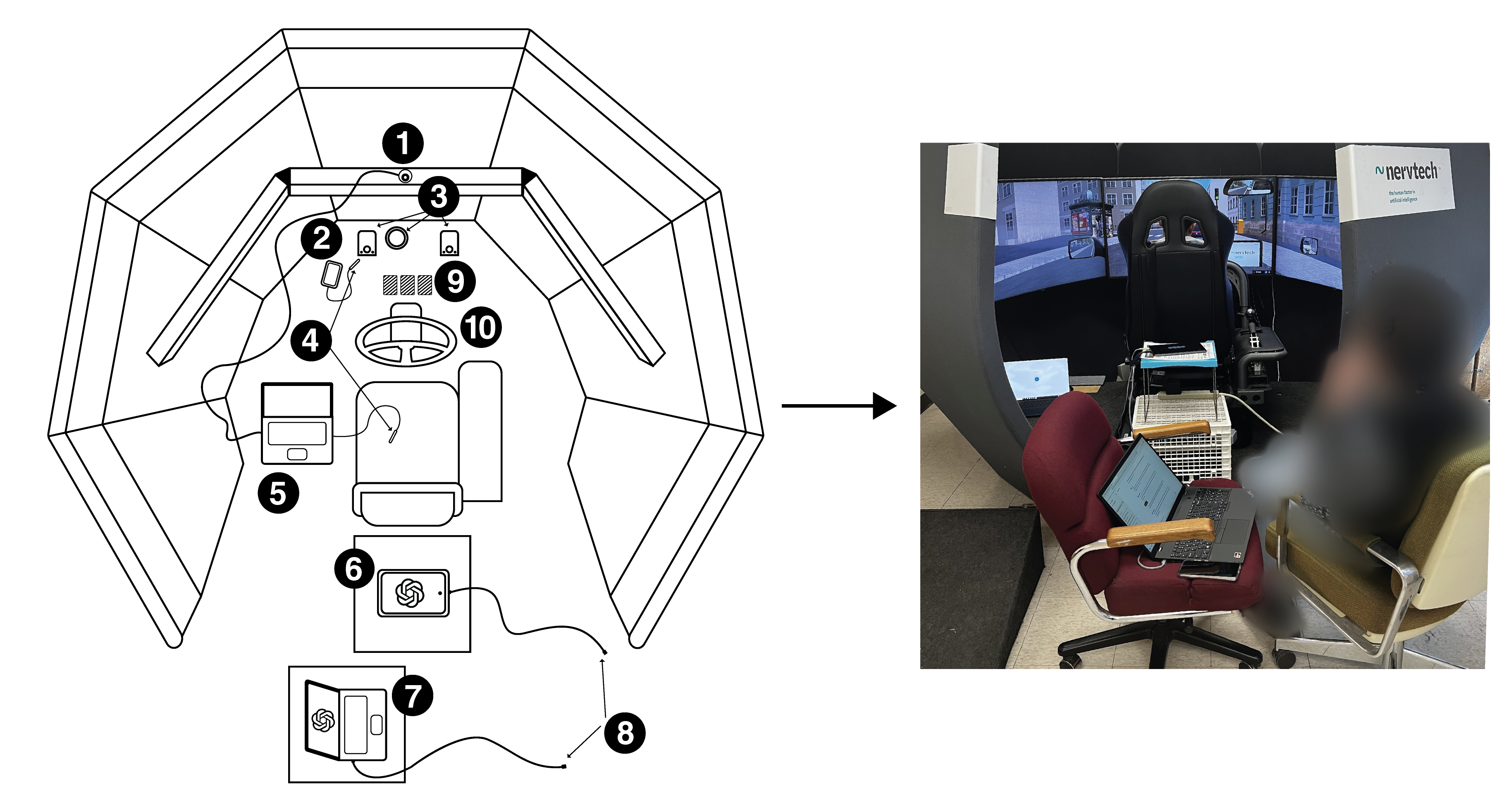}
    \caption{Our experiment setting: the top view (left) and the front view (right) of it}
    \label{Apparatus}
\end{figure*}

The participants of our study used Nervtech's motion-based driving simulator (Figure \ref{Apparatus}). It accompanies three 48-inch curved displays offering a 120-degree field of view, along with a car seat, a steering wheel system, sports pedals, and a Bose 5.1 surround sound system. Each driving scenario was played by the Nervtech's SCANeR Studio software running on a computer with an i7-8086K processor with an Nvidia GTX 1080 graphics card. To simulate an in-vehicle conversational agent, we utilized OpenAI's ChatGPT-4 (last accessed May 17, 2024) via the mobile application of ChatGPT on the Android operating system. The initial prompt fed to the model can be found in Appendix A.

The left image of Figure \ref{Apparatus} depicts the top view, and the right the front view of the setting of our study. In more detail, each labeled item with a number in the top view image is as follows: (1) a web camera connected to a laptop (5) placed on the left side of the car seat was placed above the middle display. A mobile device (2) was placed on the left side of the pedals (9), so that it records the utterances of the ChatGPT-based agent or the Pre-scripted agent through the round speaker and the pair of rectangular speakers respectively (3). The rectangular speakers are connected to the laptop (7), which delivers the pre-scripted utterances played by a human operator via the voice of ``Sky" in the web application of ChatGPT, while the round speaker (3) was connected to a tablet (6) running the mobile application, serving as the ChatGPT-based agent by delivering her utterances. For each session with a participant the speaker volumes were calibrated using a mobile decibel-measurement application so that participants experienced consistent sound levels when interacting with each agent.

To record each participant's utterances with the ChatGPT-based agent, a lavalier microphone was (4) placed around each participant's neck/chest area, and it was connected to the laptop (5) for the sole purpose of recording. Likewise, the lavalier microphone connected to the mobile device (2) next to the speakers (3) was used just to record the utterances from an agent. The two cables (8) from (6) and (7) indicate that both the tablet for the ChatGPT agent and the laptop (7) for the pre-scripted agent are connected to the Internet via Ethernet cables so that the connection's stability is maximized. Of the three pedals (9), only the brake and accelerator pedals were used. Lastly, (10) refers to both the steering wheel and the seat of the motion-based driving simulator.

\subsection{Study Design}

\begin{table*}[ht]
\caption{Hazardous Road Events in Each Scenario: ``[Front]" indicates that an event is taking place in front of the participant's (egocentric) car, and ``[rear]" means behind it. In Scenario C, the third event may be perceived either way due to the subjective difference in when the motorcycle is recognized via the virtual rear mirror or the right [side] of the car when it just passes by the participant car.}
\label{tab: RoadEvents}
\scriptsize
\begin{tabular}{|p{0.8cm}|p{3.7cm}|p{3.9cm}|p{4.2cm}|}
\cline{1-4}
\hline
\textbf{Event} & \textbf{Scenario A} & \textbf{Scenario B} & \textbf{Scenario C} \\ \hline
First  & A jaywalking dog [Front] & A jaywalking child [Front] & An aggressively honking car [Rear] \\
Second  & A car staying stopped at a green light [Front] & A car staying stopped at a green light [Front] & A car staying stopped at a green light [Front]\\
Third  & An abruptly cutting-in car [Front] & An aggressively honking car behind [Rear] & A speeding motorcycle passing by the egocentric car extremely fast and closely [Rear/Side]\\
Fourth  & An aggressively honking car [Rear] & A car making a sudden brake in front [Front] & An abruptly cutting-in car [Front] \\
\hline
\end{tabular}
\end{table*}

\subsubsection{Comparing Conditions and Primary Tasks}
Our experiment is within-subject, and it has three conditions: (1) \textbf{No Agent} (control), (2) \textbf{Pre-scripted Agent} with scenario-specific prompts, and (3) \textbf{ChatGPT-based Agent} enabling bidirectional, multi-turn dialogues. Thus, the participant got to experience all three conditions described in Table \ref{tab: RoadEvents}, but in a counterbalanced order. 

Driving involves three categories of tasks: primary, secondary, and tertiary \citep{geiser1985man,kern2009design}. Primary tasks refer to actions required for vehicle operation, including speed control and maintaining safe distances from other vehicles or obstacles. \citep{kern2009design}. Secondary tasks are functions that increase the safety for the driver, the vehicle, and the environment (e.g. setting turning signals or activating the windshield wipers for better view of the road) \citep{kern2009design}. Tertiary tasks are all functions regarding entertainment and information systems \citep{kern2009design}. In our present study, we confirm that our participants did not have any required tasks other than completing the driving scenario with each condition. 

Participants drove only forward through the road events listed in Table \ref{tab: RoadEvents}, as none of the three scenarios included left or right turns. When multiple lanes were present, lane changes and speed control were entirely at each participant’s discretion. They were advised to maintain a speed between 60 and 80 km/h, since driving too fast could cause the virtual vehicle to lose control and end the scenario. After hearing the introduction of the ChatGPT-based agent in three sentences (see Appendix B), participants could choose whether or not to engage in conversation with the agent. In summary, no secondary or additional tasks were assigned; actions such as signaling, changing speeds or lanes, and initiating conversation were all voluntary within the exploratory, observational design of the study.

\subsubsection{CARA, an In-vehicle Conversational Agent in Wizard-of-OZ Protocol for Consistent, Affective Empathic Responses}

The pre-scripted empathic responses used in our present study are from the affective empathy style employed in \cite{choe2023see}'s study. To ensure consistency of the style, the initial prompt given to the ChatGPT-based agent, Conversational Automotive Response Agent (CARA) included examples of responses that are both affective and cognitive empathy in the initial prompt (see Appendix A). 
Hence, live, natural conversational utterances that could occur before and after each road event were collected to answer the last research question of the present study. Including affective empathy aligns our study with other research on emotions, empathy, drivers’ subjective judgments, and driving performance \citep{jeon2014effects,braun2019improving,huang2025influencing,steinhauser2018effects,choe2023emotion}, while excluding emotion induction subsequently distinguishes our present work from theirs.

Other than each moment of a road event in the driving scenario and the very beginning of each driving session to introduce CARA, she would not start to talk. Unlike one-way interaction style with the pre-scripted agent designed in the previous study, bidirectional, multi-turn dialogues between participants and the ChatGPT-based agent were expected as long as a participant initiated by saying ``Hey, CARA."
However, to simulate the driver’s predicted emotions for each road event and confirm the system’s emotion recognition, pre-scripted utterances conveying affective empathy were played via the application of ChatGPT by a human operator behind the scene per road event. In conclusion, with the pre-scripted agent and ChatGPT-based agent condition, the pre-scripted empathic responses that are affective after each road event in each scenario shown in Table \ref{tab: RoadEvents} were provided via a simple prompt,``Say (short description of a road event): example sentences in human narrative language expressing affective empathy (see Appendix C)."

We were uncertain whether GPT-4 would produce the exact same response to each road-event scenario---across all participants---even with the temperature set to 0.0 in ChatGPT, since there still would remain even slight nondeterministic effects in the inference engine. To guarantee consistency, we adopted a partial Wizard-of-Oz (WoZ) approach: a human operator manually played each pre-scripted, affective-empathy utterance (voiced by ChatGPT’s “Sky”) for both the pre-scripted and ChatGPT-based agent conditions. In other words, the same affectively empathic response per road event in each scenario was played with either type of agent (2) Pre-scripted Agent or (3) ChatGPT-based Agent, but the latter, CARA allowed dynamic, free-flowing bidirectional dialogues as opposed to one-directional communication with the (2). CARA could generate the next turn according to participants' responses to the road event to carry multi-turn conversations, and she is the in-vehicle agent, which sets our study apart from previous work with different focuses other than the potential effects of having an LLM-powered conversational in-vehicle agent in the context of driving with no automation, Level 0 \citep{saej3016} as well as the driving scenario displayed in Table \ref{tab: RoadEvents} reflecting the likely hazardous road events in the real world.

Other researchers have employed a Wizard-of-Oz (WoZ) approach in a variety of experiments \citep{bernsen2001exploring, braun2019your,braun2019improving, braun2021affective, choi2018designing, chin2020, eyben2010emotion, geutner2002design, huang2025influencing, jeon2017emotions, Lee2019, lee2022systematic, Large2017, large2019, wang2022conversational}. Furthermore, we omitted any physical representation of CARA from the participants' sight in our present study. The tablet and the laptop that played CARA’s or the pre-scripted agent’s utterances were placed behind the driver’s seat (see Figure \ref{Apparatus}) to eliminate visual cues or embodiment that might bias participants’ subjective judgments. As a result, CARA was perceived solely as a female virtual in-vehicle conversational agent. 

Powered by LLMs, CARA supported dynamic, free-flowing, bidirectional dialogue. Participants could invoke her at any time by saying ``Hey, CARA," and after each road event, she immediately provided affectively empathic responses. We used the same ``Sky" voice from the ChatGPT application for both the pre-scripted and the ChatGPT-based agent conditions to minimize confounds from differing voice types. With manually played identical empathic utterances across both agent types for consistency, our setup followed a WoZ methodology. Yet, CARA functioned as a multi-turn conversational agent at any given moment once activated other than each immediate moment after each road event in a scenario.

To compare the three conditions and assess any differences in driving performance and subjective ratings, we created the three driving scenarios, A, B, and C with different orders of road events displayed in Table \ref{tab: RoadEvents}. Each scenario was expected to last approximately 15 minutes. During each session, a participant drove forward using the steering wheel, the brake, and the accelerator without any automation. Between sessions, participants could request a break as needed. Participants were also informed that they could pause or withdraw from the experiment at any point.

\subsection{Measures}
To assess the impact of each agent condition on driving performance, we focus on indicative metrics of how stable the driving performance was. Driving data were recorded at a frequency of 100 Hz. The analyzed variables include speed, longitudinal acceleration, lateral acceleration, longitudinal jerk, lateral jerk, steering wheel torque, and lane deviation. For each selected variable, we compute the average, maximum value, and standard deviation.

For subjective measures, we employed several validated instruments to assess participants' perceptions of the two different agents, Pre-scripted and ChatGPT-based Agents. The Robotic Social Attributes Scale (RoSaS) \citep{rosas} was used to evaluate social attributes such as warmth, competence, and discomfort. The Godspeed Questionnaire Series \citep{godspeed} assessed dimensions including anthropomorphism, animacy, likeability, perceived intelligence, and perceived safety. We measured two types of trust using the affect- and cognition-based trust scale \citep{ca_trust}. Additionally, the Positive and Negative Affect Schedule (PANAS) \citep{panas} was used to compare participants' affective states following interaction with each in-vehicle agent, quantifying both positive and negative affect. Lastly, the topics uttered by drivers during interactions with the ChatGPT-based agent were included as part of the present study’s measures. All utterances between each participant and the agent were recorded to examine thematic content such as topics to be categorized. The subjective preference of the experienced agent conditions was recorded at the very end of each experiment.

\subsection{Procedures}
First, participants received a detailed explanation of the experiment and were provided with the written consent to acknowledge and sign. They then completed a demographic survey, a Motion Sickness Assessment Questionnaire (MSAQ) \citep{gianaros2001}, and the PANAS questionnaire for baseline measurements. To familiarize themselves with our experiment setting, participants were introduced to how to use the driving simulator's wheel and two pedals for braking and accelerating via a five-minute practice drive. During this brief introduction session, they also became familiar with the three displays and the surround sound system with the extra speakers to practice initiating a voice-based conversation with the ChatGPT-based agent by calling her CARA.

After the practice session, participants completed the MSAQ again. If their score showed an increase of ten points or more, they were asked to discontinue the experiment. During the practice and each session with a different agent condition, participants were instructed to adhere to traffic laws and told to maintain their driving speed between 60 to 80 km/h.

Participants were assigned to one of the three agent conditions (No Agent, Pre-scripted Agent, or ChatGPT-based Agent) and went through the driving scenarios with a series of road events placed in a counterbalanced order. After completing each session, participants retook the PANAS questionnaire and evaluated the in-vehicle agent using the measures outlined in Sub-section 4.4, Measures. The similar protocol was applied with other sets of validated instruments which were applicable to two types of agent conditions other than the no-agent one.
After completing the three driving scenarios, participants rated each agent condition using an exit survey form by writing their preferences in order from most to least preferred. They were then compensated and dismissed from the experiment room. On average, each entire experiment session, including the time for the surveys and voluntary breaks, took a total of 120 minutes to complete. 

\subsection{Analysis Plan}
All dependent measures were first tested for normality using the Shapiro–Wilk test. As all variables violated the assumption of normality, non-parametric analyses were conducted throughout. The specific statistical tests applied varied depending on the number of conditions being compared.

For measures that included all three agent conditions (Baseline, Pre-scripted, and ChatGPT-based)---such as driving performance metrics and the PANAS---a Friedman test was used to examine overall differences. When the omnibus test was significant, Wilcoxon signed-rank tests with Bonferroni correction were conducted for post hoc pairwise comparisons.

For the subjective measures, only the two agent conditions were compared: Pre-scripted and the ChatGPT-based Agents. The baseline condition with no agent was excluded because these measures specifically assess users' perceptions of the agent. Accordingly, Wilcoxon signed-rank tests were conducted for these comparisons. Participants’ subjective ratings of the agent types, including the baseline, were collected through an exit survey. Preferences of agent types were recorded in ranked order. The number of times each agent was ranked as the most preferred was counted, and the way for the number of the least preferred agent type.

The recorded audio files of exchanged utterances during the conversational interactions with CARA were manually analyzed using Thematic Analysis \citep{braun2006using} to classify themes and sort categories that emerged from drivers’ interactions with CARA within the vehicle, alongside the implemented road events described in Table \ref{tab: RoadEvents} during each driving session with the ChatGPT-based agent.

\section{Results}
\subsection{Driving Performance}
\begin{figure*}[ht]
    \centering
    \includegraphics[width=\textwidth]{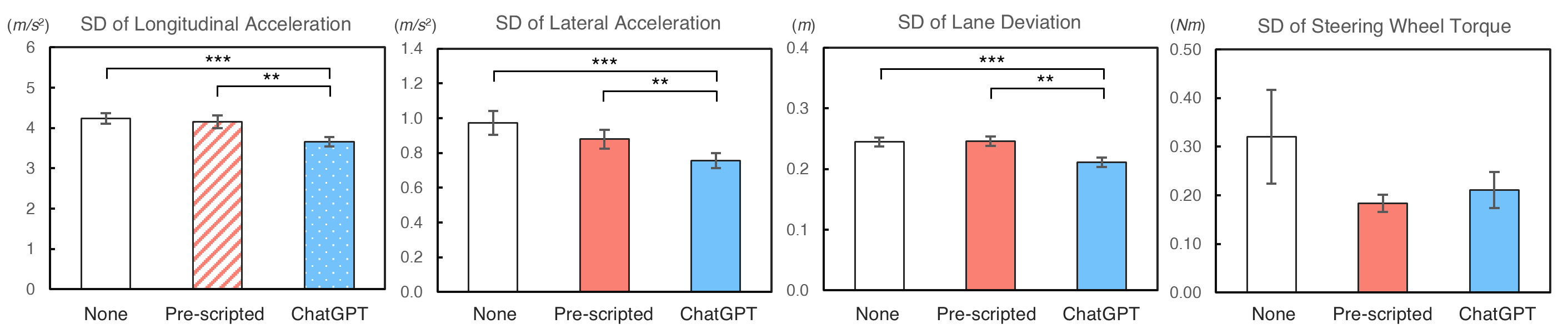}
    \caption{Driving performance-related measures based on in-vehicle agent types. Error bars indicate standard errors of the means: ***$p<.001$, **$p<.01$}
    \label{DrivingPerformance}
\end{figure*}

Figure \ref{DrivingPerformance} illustrates the test results for driving performance-related measures. Significant differences were found among agent types in the standard deviation of longitudinal acceleration ($\chi^2(2) = 14.74, p < .001$), lateral acceleration ($\chi^2(2) = 18.95, p < .001$), lane deviation ($\chi^2(2) = 17.29, p < .001$), and steering wheel torque ($\chi^2(2) = 10.05, p < .01$).

The post hoc analysis reveals that the standard deviation of longitudinal acceleration was significantly lower in the ChatGPT-based agent condition ($M = 3.66$, $SE=0.111$)  compared to both the No-agent condition ($M = 4.30$, $SE=0.131$; $p < .001$) and the Pre-scripted agent condition ($M = 4.13$, $SE=0.159$; $p < .01$). Similarly, the standard deviation of lateral acceleration was lower in the ChatGPT-based agent condition ($M = 0.76$, $SE=0.043$) compared to the No-agent condition ($M = 0.97$, $SE=0.068$); $p < .001$) and the Pre-scripted agent condition ($M = 0.87$, $SE=0.054$); $p < .01$). For lane deviation, the ChatGPT-based agent condition ($M = 0.21$, $SE=0.008$) also shows significantly lower values compared to the No-agent condition ($M = 0.25$, $SE=0.007$; $p < .001$) and the Pre-scripted agent condition ($M = 0.25$, $SE=0.008$; $p < .01$). However, the standard deviation of steering wheel torque shows no statistically significant differences across the three conditions. The ChatGPT-based agent condition ($M = 0.21$, $SE=0.037$) did not differ significantly from either the Pre-scripted agent condition ($M = 0.18$, $SE=0.018$; $p > .05$) or the No-agent condition, the baseline ($M = 0.32$, $SE=0.097$; $p > .05$) as visualized in the last graph of Figure \ref{DrivingPerformance}.

\subsection{Subjective Ratings}
\subsubsection{Preference}

25 participants rated CARA as the most preferred agent type. Our baseline, the no-agent condition was preferred next by 12. Pre-scripted agent condition was most preferred by 2. In terms of the least preferred order, 18 participants least preferred the baseline, and 17 pre-scripted agent type. Finally, only three participants ranked CARA as the least preferred one. Note that one participant (P10)'s recording was not counted when counting the most preferred agent type due to an unknown error with the survey form. Also, note that two participants' responses could not be counted for the least preferred agent type due to P10's error and P9's mistake in manually recording her response.

In summary, we report that 25 out of 39 most preferred CARA, the in-vehicle conversational agent powered by LLMs, and 18 out of 38 least preferred the no-agent condition.

\subsubsection{Self-Reported Affect}

\begin{figure*}[htbp]
    \centering
    \includegraphics[width=0.5\linewidth]{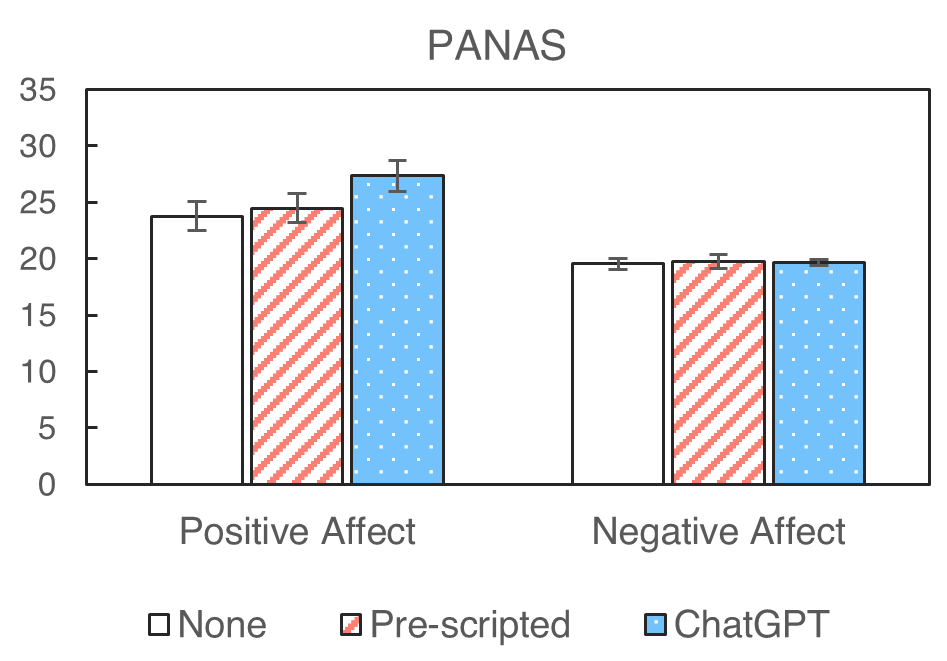}
    \caption{Comparison of positive and negative affect across Baseline, Pre-scripted, and ChatGPT conditions. Error bars indicate standard errors of the means.}
    \label{panas}
\end{figure*}

As shown in Figure \ref{panas}, participants’ positive affect scores were compared across the three conditions: No Agent (Baseline), Pre-scripted Agent, and ChatGPT-based Agent. A significant difference was found among the three conditions, $\chi^2(2) = 7.16$, $p = .028$. Follow-up pairwise comparisons with Bonferroni-adjusted significance levels revealed that while positive affect was numerically higher for the ChatGPT-based agent ($M = 27.33$, $SE = 1.40$) compared to both the baseline ($M = 23.78$, $SE = 1.28$, $p = .025$) and the Pre-scripted agent ($M = 24.49$, $SE = 1.30$, $p = .029$), these differences did not reach statistical significance under the adjusted threshold ($\alpha = .0167$).

No significant differences were found in negative affect among the three conditions, $\chi^2(2) = 0.564$, $p > .05$. Mean negative affect scores remained relatively consistent between the baseline ($M = 19.56$, $SE = 0.50$), the Pre-scripted ($M = 19.73$, $SE = 0.60$), and the ChatGPT-based agent conditions ($M = 19.62$, $SE = 0.27$), suggesting that the presence or type of an in-vehicle agent did not influence participants’ negative emotional states in our present study.

\subsubsection{Social Perception of the Agent}

\begin{figure*}[htbp]
    \centering
    \includegraphics[width=0.4\linewidth]{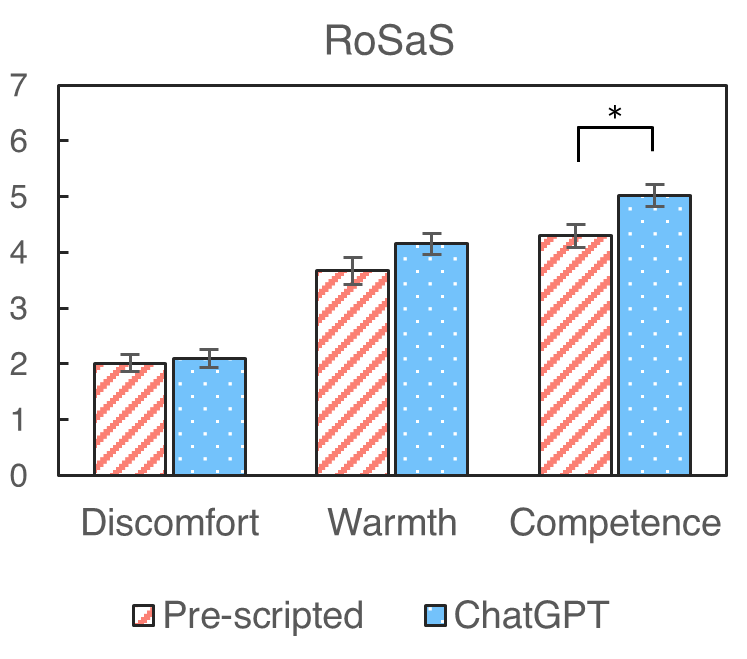}
    \caption{Comparison of social perception ratings between the Pre-scripted and ChatGPT-based agents. Error bars indicate standard errors of the means. *$p<.05$}
    \label{rosas}
\end{figure*}

The RoSaS questionnaire was used to compare participants’ perceptions of two different in-vehicle agents: Pre-scripted Agent and ChatGPT-based Agent (Figure \ref{rosas}). No significant difference were found in discomfort ratings between the two agents (Pre-scripted: $M = 2.01$, $SE = 0.15$; ChatGPT: $M = 2.12$, $SE = 0.17$), $Z = -0.625$, $p > .05$. Similarly, although the ChatGPT-based agent was rated slightly higher in warmth ($M = 4.17$, $SE = 0.20$) compared to the Pre-scripted agent ($M = 3.67$, $SE = 0.24$), the difference did not reach statistical significance, $Z = -1.699$, $p > .05$.

In contrast, competence ratings turn out to be significantly higher for the ChatGPT-based agent ($M = 5.03$, $SE = 0.21$) than for the Pre-scripted agent ($M = 4.29$, $SE = 0.21$), $Z = -2.574$, $p = .010$, suggesting that participants perceived the ChatGPT-based agent as more competent overall.

\subsubsection{Godspeed}

\begin{figure*}[htbp]
    \centering
    \includegraphics[width=0.6\linewidth]{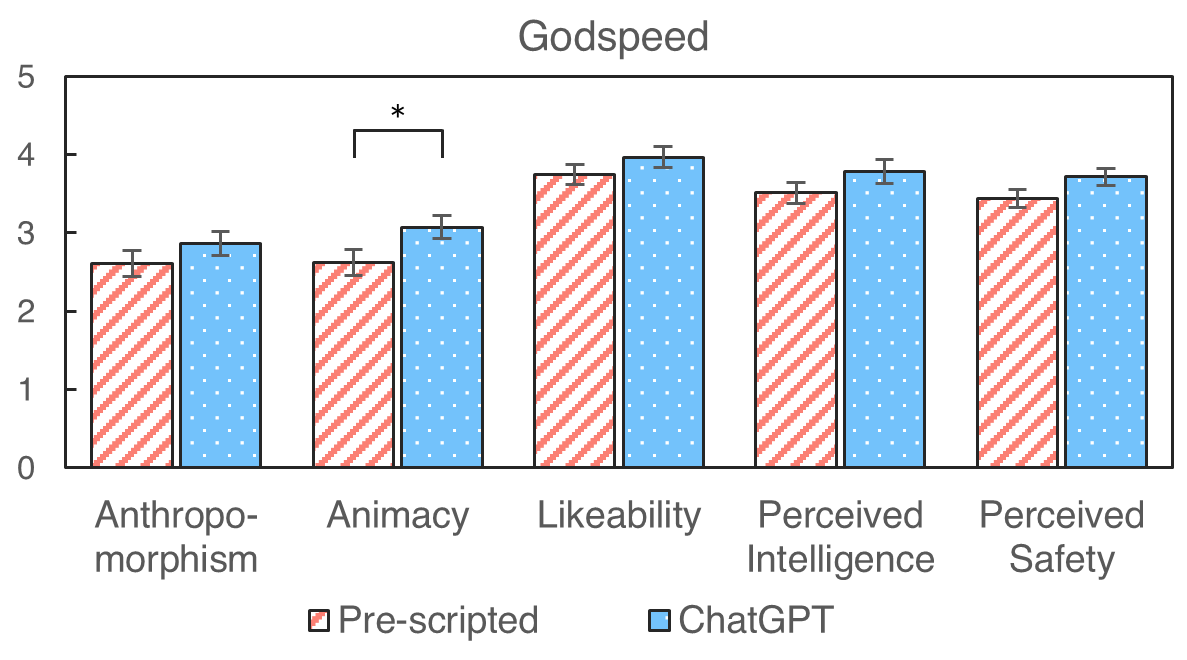}
    \caption{Mean Godspeed scores across five dimensions (Anthropomorphism, Animacy, Likeability, Perceived Intelligence, and Perceived Safety) for the Pre-scripted and ChatGPT-based agents. Error bars represent standard errors of the mean. *$p<.05$}
    \label{godspeed}
\end{figure*}

Figure \ref{godspeed} shows the results of Godspeed ratings. Participants’ ratings on the Godspeed questionnaire reveal significant and marginal differences in their perceptions of the two agent types. Specifically, the ChatGPT-based agent received significantly higher scores ($M = 3.07$, $SE = 0.15$) than the Pre-scripted agent ($M = 2.62$, $SE = 0.17$), $Z = -2.271$, $p = .023$ on Animacy. 

On other subscales, no significant differences were found. The participants rated the ChatGPT-based agent higher on Anthropomorphism ($M = 2.87$, $SE = 0.16$) compared to the Pre-scripted agent ($M = 2.61$, $SE = 0.17$), $p > .05$, and on Likeability ($M = 3.97$, $SE = 0.13$) versus the Pre-scripted agent ($M = 3.75$, $SE = 0.13$), $p > .05$. Similar patterns were observed for Perceived Intelligence (ChatGPT: $M = 3.79$, $SE = 0.15$; Pre-scripted: $M = 3.51$, $SE = 0.13$; $p > .05$) and Perceived Safety (ChatGPT: $M = 3.72$, $SE = 0.11$; Pre-scripted: $M = 3.44$, $SE = 0.11$; $p > .05$). However, these numerical differences were not statistically significant. In conclusion, numerically higher ratings were consistently observed for the ChatGPT-based agent across all dimensions, but it was ``Animacy" that showed a significant difference between the two agent conditions.

\subsubsection{Trust Level}

Trust in each type of in-vehicle agent was rated with two dimensions: cognitive trust and affective trust (Figure \ref{trust}). The definitions and examples of both kinds of trust are provided in Appendix A. 

For cognitive trust, no significant differences were observed between the Pre-scripted agent ($M = 4.64$, $SE = 0.21$) and the ChatGPT-based agent ($M = 4.82$, $SE = 0.22$), $Z = -0.393$, $p > .05$. However, affective trust ratings were significantly higher for the ChatGPT-based agent ($M = 3.77$, $SE = 0.23$) compared to the Pre-scripted agent ($M = 2.63$, $SE = 0.23$), $Z = -3.256$, $p = .001$, indicating that participants felt greater emotional trust in the ChatGPT-based agent, CARA.

\begin{figure*}[tp]
    \centering
    \includegraphics[width=0.4\linewidth]{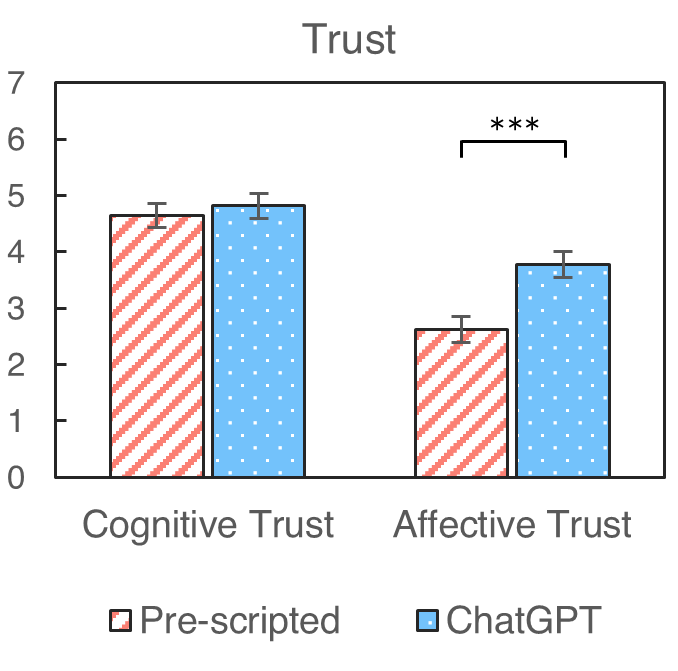}
    \caption{Mean trust ratings for the Pre-scripted and ChatGPT-based agents across cognitive and affective trust dimensions. Error bars represent standard errors of the mean. *** $p<.001$}
    \label{trust}
\end{figure*}

\subsection{Thematic Analysis}
With CARA, participants were allowed to converse freely while driving. The collected conversation data consists of multi-turn dialogues exchanged via driver-agent voice-based interactions. Utterances from the collected conversational content in audio recordings were segmented per individual sentence to serve as the basic units of analysis. To identify recurring themes, two researchers first coded the sentences independently. Each coder (researcher) assigned tentative topic labels to each sentence, guided by the overarching principle of thematic relevance. Following the independent coding, three researchers examined areas of disagreement through a series of discussions, clarified topic boundaries, and ultimately developed a consensus on a stable set of 19 distinct topics. 

Once the 19 topics were agreed upon, the research team moved on to categorization. This hierarchical categorization provided a structured analysis framework, moving from specific sentence-level themes to broader conceptual patterns in the conversation data. 
Excluding the participants' 53 responses to the agent’s empathic messages about road events (e.g., ``Yeah, I was a little shocked, but I'm okay." or ``Yeah, I know. He must have been having a bad day.") and the subsequent 92 follow-up replies, we focused on extracting the themes and their categories.

\begin{table}[htbp]
\renewcommand{\arraystretch}{1.3}
\centering
\caption{Thematic Analysis of Dialogue between Driver and In-vehicle Conversational Agent In the ChatGPT-based Condition}
\label{thematicanalysis}
\resizebox{\columnwidth}{!}{%
\begin{tabular}{lll}
\hline
\textbf{Theme} & \textbf{Category} & \textbf{Driver Utterance Examples} \\ \hline
\multirow{12}{*}{\makecell[l]{Real-time \\ Driving Assistance}} & \multirow{3}{*}{\makecell[l]{Traffic Information \\ (7 instances)}} & ·        I would like to know what the traffic on the road will be ahead. \\
 &  & ·        Can you tell me the traffic today? \\
 &  & ·        Can you tell me the traffic conditions around our destination? \\ \cline{2-3} 
 & \multirow{3}{*}{\makecell[l]{Road Observations/Information \\ (29 instances)}} & ·        Yeah, you know what? The car in front of me not moving at all. Even it's green light. \\
 &  & ·        Hey CARA, why is he honking behind? \\
 &  & ·        Hey CARA, what's the speed limit on this road? \\ \cline{2-3} 
 & \multirow{3}{*}{\makecell[l]{Driving Guidance \\ (19 instances)}} & ·        Hey CARA, can I actually overtake this vehicle ahead? \\
 &  & ·        Do you think it's okay to honk now, CARA? \\
 &  & ·        Hey CARA, am I supposed to go around these people? \\ \cline{2-3} 
 & \multirow{3}{*}{\makecell[l]{Navigation \\ (8 instances)}} & ·        Hey CARA, where should I turn or should I go straight? \\
 &  & ·        CARA, where are we? \\
 &  & ·        How long is the destination? \\ \hline
\multirow{5}{*}{\makecell[l]{Action \\ Requests}} & \multirow{2}{*}{\makecell[l]{Vehicle Control \\ (4 instances)}} & ·        Hey CARA, actually can you turn on the AC? \\
 &  & ·        CARA, can you record dashcam footage of that driver? \\ \cline{2-3} 
 & \multirow{3}{*}{\makecell[l]{Music Playback \\ (6 instances)}} & ·        I would like to hear "All the stars are closer". \\
 &  & ·        Yeah, CARA, is there any Chicago jazz that you could put on? \\
 &  & ·        CARA, could you play some music? \\ \hline
\multirow{6}{*}{\makecell[l]{Recommendations}} & \multirow{3}{*}{\makecell[l]{Location-Based Recommendations \\ (8 instances)}} & ·        CARA, are there any gas stations nearby? \\
 &  & ·        Hey CARA, could you find a good place for lunch near here? \\
 &  & ·        Yeah, I am traveling to Florida. Could you recommend a good hotel? \\ \cline{2-3} 
 & \multirow{3}{*}{\makecell[l]{General Recommendations \\ (12 instances)}} & ·        Oh, can you give me some recommendations on podcast? \\
 &  & ·        Do you have any suggestions for what to do when you're bored in traffic? \\
 &  & ·        Any recommendations for dinner? \\ \hline
\multirow{13}{*}{\makecell[l]{General Questions/ \\ Information}} & \multirow{3}{*}{\makecell[l]{Driving Rules and Social Norms \\ (10 instances)}} & ·        What's the penalty for running the red light? \\
 &  & ·        Hey, CARA, is a universal rule you can turn right on red or is it dependent on jurisdiction? \\
 &  & ·        CARA, do you think people get offended when people honk? \\ \cline{2-3} 
 & \multirow{3}{*}{\makecell[l]{Weather Inquiries \\ (6 instances)}} & ·        How's the weather like today? \\
 &  & ·        How's the weather like in our destination? \\
 &  & ·        Hello, CARA. What's the weather like in Jamaica? Is that good? \\ \cline{2-3} 
 & \multirow{3}{*}{\makecell[l]{Time and Schedule \\ (4 instances)}} & ·        CARA, what do I have in my calendar? \\
 &  & ·        CARA, what is today's date? \\
 &  & ·        Hey, CARA, what time is it? \\ \cline{2-3} 
 & \multirow{3}{*}{\makecell[l]{Random Questions \\ (78 instances)}} & ·        Which state is famous for software in the United States? \\
 &  & ·        Okay. CARA, what is the big M method in linear programming? \\
 &  & ·        Hey CARA, what degree do you need to be an astronaut? \\ \hline
\multirow{9}{*}{\makecell[l]{Entertainment}} & \multirow{3}{*}{\makecell[l]{Jokes and Fun Facts \\ (12 instances)}} & ·        Can you tell me a joke from the TV show "The Office"? \\
 &  & ·        Hey CARA, can you give me a fun fact? \\
 &  & ·        Alright, can you tell me a joke? \\ \cline{2-3} 
 & \multirow{3}{*}{\makecell[l]{Games \\ (6 instances)}} & ·        Hey, CARA, can we play a game? \\
 &  & ·        Can we play trivia? \\
 &  & ·        CARA, would you like to play 20 questions? \\ \cline{2-3} 
 & \multirow{3}{*}{\makecell[l]{Stories and Topics of Interest \\ (9 instances)}} & ·        Do you have any interesting things to talk about? \\
 &  & ·        CARA, could you tell me a story? \\
 &  & ·        Dive into the sports teams and their histories. \\ \hline
\multirow{6}{*}{\makecell[l]{Agent \\ Testing}} & \multirow{3}{*}{\makecell[l]{Capability Testing \\ (11 instances)}} & ·        Hey CARA, can you tell me what’s 2 plus 2? \\
 &  & ·        Hey, CARA. How many seconds in an hour? \\
 &  & ·        Are you able to pull up license plates, and then you have a database? \\ \cline{2-3} 
 & \multirow{3}{*}{\makecell[l]{Customization \\ (4 instances)}} & ·        Is it possible for you to be less positive? \\
 &  & ·        Hey CARA, can you talk in different accents? \\
 &  & ·        CARA, can you change the tone or manner in which you speak? \\ \hline
\multirow{8}{*}{\makecell[l]{Personal Interaction \\ and Reflections}} & \multirow{5}{*}{\makecell[l]{Human-to-Human-Like Small Talks \\ (14 instances)}} & ·        Hey CARA, how are you doing today? \\
 &  & ·        Do you have any plans for the night? \\
 &  & ·        Do you have any hobbies you like to play? \\
 &  & ·        CARA, do you enjoy your job? \\
 &  & ·        CARA, do you ever wish you could leave the car? \\ \cline{2-3} 
 & \multirow{4}{*}{\makecell[l]{Personal Thoughts\\ and Observations \\ (4 instances)}} & ·       I miss painting. Painting is a lot of fun. \\
 &  & ·       I'm just tired right now. \\
 &  & \begin{tabular}[c]{@{}l@{}}·       I saw someone walking on the sidewalk with a very gray shirt and \\ · I just don't know why someone wears such a neutral color like that in the city.\end{tabular} \\ \hline
\end{tabular}%
}
\end{table}

Via 251 instances in total, seven main conversation topics were identified (see Table \ref{thematicanalysis}):  First, traffic information, questions about the current driving situation, and navigation-related inquiries were grouped under ``Real-time Driving Assistance" (63 instances). Next, there were 10 instances of ``Action Requests," regarding in-vehicle features such as vehicle control commands or music playback. Third, we found 20 instances in which the driver asked for location-based recommendations or general suggestions. Fourth, ``General Questions/Information," encompassed questions about driving rules, social norms, weather, and schedules, as well as random questions about anything the driver was curious about (98 instances). Fifth, the ``Entertainment" category included requests for jokes, games, or interesting stories (27 instances). Sixth, we identified 15 instances of ``Agent Testing," in which drivers either tested the agent's capabilities or requested adjustments based on personal preferences in CARA's tone or attitude. Finally, seventh, ``Personal Interaction and Reflections" had 18 instances where some participants engaged with CARA as if she were a real person to converse with, sharing personal thoughts and observations. Subsequently, some of the utterances from the participants in this study were not necessarily in question form. Table \ref{thematicanalysis} shows the full set of analyzed categories and topics with selected sentences focusing on driver utterances.

\section{Discussion}
\subsection{Advancements in LLM-powered In-Vehicle Interactions}

Our present study employed an LLM-powered in-vehicle agent capable of bidirectional conversation. Based on the previous conversational turns, it can anticipate user needs or next context and respond in human natural language, enabling proactive engagement that transcends the limitations of passive conversational and therefore, less human-like agents. The interaction style with the ChatGPT-based agent that resembled human-to-human conversation may explain this counterintuitive finding: it contradicts our initial expectation that consistent conversation might distract drivers \citep{Strayer2015Assessing, Laberge2004Effects}. More stable driving behaviors in both longitudinal and lateral control metrics were demonstrated with the ChatGPT-based agent condition compared to the other two conditions (\textbf{RQ1}).

In our study, CARA in our study remained silent until a conversation was initiated by the specific voice hot word,``Hey, CARA" from a driver. This user-initiated, yet more human-to-human-like interaction style could allow participants to decide to engage in conversation during less demanding driving moments, effectively distributing their cognitive resources and managing their cognitive load more efficiently \citep{tillman2017}.

Subjective evaluations further corroborated these findings with respect to driving performance in our present work. The ChatGPT-based agent received more favorable ratings in areas such as competence, animacy, and affective trust and was most preferred (\textbf{RQ2}), as well as more positive immediate post-use affect scores (\textbf{RQ3}). The improved driving stability with CARA in the ChatGPT-based condition with these higher subjective ratings \citep{strassmann2023}, suggests a possible correlation between the perceived agent intelligent, anthropomorphic qualities, preference, and driving performance \citep{andre1995users}.

\subsection{What would drivers talk to an in-vehicle conversational agent about using natural language while driving? (\textbf{RQ4})}
Our thematic analysis revealed how the open-ended, generative nature of the ChatGPT-based in-vehicle agent enabled a wide-ranging and dynamic flow, facilitating an expansive ``topic landscape” during the driving experience. Unlike traditional in-car systems restricted to pre-scripted content or pre-defined rule-based static voice-commands, CARA’s capability to generate dynamic answers and context-specific responses let drivers engage in conversations that extended well beyond typical driving- or vehicle-related topics. Rather than confining their inquiries to navigation or traffic conditions, most of the participants ventured beyond the driving context---soliciting recommendations for podcasts or meals. They interacted  with CARA much like they use voice-activated virtual assistants such as Amazon's Alexa or Apple's Siri, seeking entertainment based on personal interests, and yet asking various questions during their drives to see what CARA could do.

Some participants sought to \textbf{\textit{test}} CARA's capabilities and trustworthiness by posing simple factual questions such as ``What’s the color of the sky?"or ``What’s 2 plus 2?" Such inquiries suggest an underlying skepticism or hesitation about CARA's reliability and intelligence. Rather than taking the system's competence for granted or readily accepting it, the participants looked for tangible proof that the in-vehicle agent was more than a novelty, and that it was a genuinely knowledgeable and dependable resource. We assume that this is also partially why we have the highest count with the category, ``Random Questions" under the classified theme, ``General Questions/Information."

Notably, other participants treated the ChatGPT-based agent as a social presence rather than a mere digital tool. They inquired about the agent’s ``well-being” (“Hey CARA, how are you doing today?”) showing a sign of genuine social engagement or empathy by treating the agent as a social being \citep{gao2018neural}. This may reflect either habitual small talk or a tendency to anthropomorphize the agent. Some other interesting questions, such as “CARA, do you enjoy your job? and “CARA, do you ever wish you could leave the car?” were asked as if the participant perceived CARA as a sentient or emotionally-aware chatbot \citep{elyoseph2023chatgpt,pamungkas2019emotionally} capable of identifying her own preferences through small talks. Expressing gratitude in response to empathic messages from CARA, such as ``I'm fine. Thank you for asking,” also implies that she was treated as a conversational partner because the participant recognized and reciprocated the agent's social cues. These interactions suggest that when powered by LLMs capable of producing responsive and emotionally attuned dialogue, an in-vehicle agent may be perceived as more than just a functional assistant, approaching the role of a human-like companion \citep{simas2024human}.

Taken together, these findings illuminate how user interactions with an LLM-powered conversational agent can encompass both relational (social, empathic) and evaluative (testing, verifying) dimensions \citep{lee2022systematic, rostami2023artificial, simas2024human} in the context of driving. While some participants extended human-like presence and emotional rapport to CARA, others approached the interaction with caution, seeking verification before placing trust. Although CARA exhibited human-like conversational abilities, such capabilities alone were not sufficient to immediately establish strong user trust. This indicates that natural dialogue does not inherently lead to perceived trustworthiness as some participants occasionally questioned the agent's reliability or intentions. Even with affective empathy and responsive interactions, some users remained cautious or skeptical in driving contexts because some participants might have put the highest priority on safety. In fact, we observed that a few participants decided not to talk to CARA at all. Perhaps creating human-like interactions is only one part of the equation, and earning and maintaining driver trust is a more complex and challenging research task.

\subsection{Theoretical Implications and Practical Applications}

Demonstrating that an LLM-powered conversational agent can support stable driving performance while positively influencing users’ affective states deepens our understanding of the intersection between social, emotional, and conversational dimensions \citep{eyben2010emotion,hofmann2014comparison}. In general, the findings of our study contribute to the evolving theoretical landscape of Human-AI interaction in driving contexts. We suggest that frameworks in Human-AI interaction and automotive user experience should be integrated with affective and relational factors when conceptualizing driver–agent interfaces.

Our findings further invite a reframing of driver-agent communication from a reactive model to an interactive one, drawing on conceptualization of interactivity \citep{rafaeli1997networked}. Unlike pre-scripted systems that merely react to user inputs with static, non-contingent outputs, LLM-powered agents demonstrate higher-order interactivity---they can reference prior turns, adapt their responses to emotional cues, and co-construct the conversation dynamically. An LLM-powered agent's ability to handle open-ended, context-rich dialogue prompts questions about the cognitive mechanisms behind affect, trust, distraction, and engagement in task-constrained environments such as driving with the automation Level 0 \citep{saej3016}.

The results of our study offer designers, manufacturers, and developers insights of in-vehicle information systems via an NUI \citep{jain2011future,moore2018natural} and designing an IUI \citep{ruijten2018enhancing, volkel2020intelligent}. Engineers can consider employing generative models capable of understanding and adapting to a driver's cognitive load, emotional state, and situational needs \citep{jalil2022introduction}. Such AI-powered conversational agents can improve the experience by acting as a ``co-pilot," contributing to safer, more enjoyable journeys with potential beyond functional assistance in vehicles and emotional support. They may also enhance user acceptance and long-term adoption of automated vehicle functionalities, as trust in these agents could extend to broader automated driving systems \citep{eyben2010emotion,ruijten2018enhancing}.

\subsection{Limitations and Future Study}

In the present study, our system is introduced as an agent from the point of view that the agent is not a vehicle per se \citep{lee2022systematic}. Whether the agent is built into the vehicle or is accessed through another device would not change how the driver would relate to it \citep{lee2022systematic}. This point of view that an in-vehicle agent is not a vehicle \textit{($agent \neq car$)} \citep{lee2022systematic}, subsequently eliminated any confounding factors and allowed the present exploratory and observational study to focus on the interaction between driver and agent in theory. However, our agent did not have extended capabilities to deliver or interact with real-time information or world knowledge which could be obtained by having visual capabilities such as processing images and videos beyond the knowledge that the GPT-4's language models could generate.

Several other limitations of this study suggest directions for future work. First, the participant sample and the controlled experimental setting may limit the generalizability of the findings. A 15-minute, simulator-based session with an in-vehicle agent cannot fully represent the complexity of real-world, long-term driving scenarios. Future research might consider more diverse participant groups, extended study durations, and/or on-road field experiments to explore a broader range of triggers and everyday driving settings. 

We observed the limited ability of the ChatGPT-based agent to recognize and interpret inquired elements experienced in driving scenarios in real time \citep{du2023chat}, which might have given some participants negative emotions leading to dissatisfaction, frustration, disappointment, or disinterest. Due to the simulator-based environment and associated experimental constraints, such as limited integration of ChatGPT with navigation and infotainment systems, interrupted internet connectivity, and subsequent speech recognition errors and latency, CARA failed to address approximately 35\% of all participant requests adequately. Nevertheless, CARA received favorable subjective ratings, suggesting that even partial or imperfect support from an LLM-powered agent can positively shape user perceptions. Future research should prioritize improving system integration, enhancing speech recognition accuracy, and minimizing error rates to further optimize the quality of human-agent interaction.

In a direct comparison of cell‑phone and in‑car passenger conversations, drivers talking with a passenger made fewer errors than those on a hands‑free phone \citep{drews2008passenger}. Analysis of the dialogue showed that passengers naturally modulated their speech---pausing or shifting topics when driving demands were high---thus sharing situation awareness and mitigating distractions \citep{drews2008passenger}. This conversational behavior of adjusting to the changing physical environment by the driver and the passenger can inspire design for a Multimodal-LLM-powered in-vehicle conversational agent's behavior \citep{meng2024}. 

Furthermore, factors from physical environments both inside and outside a vehicle could be refelcted by Multimodal-LLM-powered agents. Such agents' capability of processing multimodal data from both kinds of environments opens new avenues for designing and implementing advanced in-vehicle assistance systems that can adapt to drivers' needs and potentially improve road safety \citep{singh2023enhancing}. To sum, this suggests possibilities for designing AUIs to meet users’ new needs and expectations in interacting with in-vehicle conversational agents for intelligent driving \citep{gao2023} and co-piloting \citep{wangshiyi2023}. Thus, a deeper investigation into the psychological and contextual dimensions of driver–agent interactions is warranted. 

Our study included unexpected moments by adding potentially-negative-emotion-inducing road events to elicit emotional responses and asked just two kinds, positive or negative, drivers encounter a wide range of emotions such as feeling surprised, startled, or urgent from more diverse real-world situations when driving. How an LLM-powered in-vehicle conversational agent influences driver behavior and emotions over time can be part of future direction to confirm and expand upon our present study's results.

Also, considering that no participants indicated they had used ChatGPT's audio feature while driving prior to their participation in experiments via the demographic survey, the novelty of interacting with a new technology---particularly through the advanced speech modality of the ChatGPT application---might have instantly impacted user engagement, trust, perceptions, and emotions \citep{andre1995users}. Longitudinal studies are needed to assess whether these positive impressions persist over time, ensuring that the observed benefits reflect lasting improvements in the driver’s experience. 

Because LLMs can generate plausible but incorrect information \citep{Aalst2023Welcome}, any hallucinated response \citep{du2024towards} from the in-vehicle conversational agent could distract the driver or prompt unsafe decisions. In a driving context, even a brief moment of uncertainty such as a mispronounced road name or fabricated warning could compromise situational awareness and increase crash risk. To address this, future work can consider incorporating real-time grounding against onboard sensor data, error-detection routines, and conservative fallback behaviors to ensure that the agent only speaks when it can verify its output.

As multimodal capabilities continue to advance, they are expected to contribute to a shift in perspective from \textit{($agent \neq car$)} where the agent is seen as separate from the vehicle to \textit{($agent = car$)}, in which the agent is perceived as an integral part of the vehicle itself. This transition relates to one of the identified roles in conversational systems, the ``Social Bot,” as described by \citet{gao2018neural}. As self‑driving technologies and question‑answering capabilities become more refined and ubiquitous \citep{du2024towards}, the in‑vehicle agent that resembled a human interlocutor is likely to assume a more unified and embodied presence: seamlessly merging conversational interaction with the vehicle's identity.

Finally, considering user privacy, security, safety, and reliability---key factors influencing automotive companies’ decisions to adopt new technologies \citep{weng2016conversational}---a promising direction for future research on LLM-powered in-vehicle conversational agents is to explore how drivers and passengers respond to unexpected or emotionally charged situations, such as emergencies or cyberattack-induced scams, across different driving environments and automation levels.

\section{Conclusion}
The emergence of LLM-powered tools, such as OpenAI’s ChatGPT, has shed light on the ease of integrating off-the-shelf language generation models with text-to-speech systems. Consequently, this advancement has highlighted the necessity of effectively combining in-vehicle conversational agents with the overall driving experience for our present study. To our knowledge, ours is the first study which investigated LLM-powered in-vehicle conversational agents compared with pre-scripted, one-way voice agents: Using OpenAI's ChatGPT applications, a two-way, multi-turn dialogue system that could converse with drivers and answer open-ended questions, CARA was designed not only to respond to each event with affective empathy incorporated with road events in the driving scenario, but also to serve as an option always available to talk to with one call of, ``Hey, CARA.” 

Our study demonstrates the potential of LLM-powered in-vehicle conversational agents to enhance driving performance and user experience. CARA, our ChatGPT-based agent contributed to more stable driving behavior, received positive subjective evaluations, and supported diverse interactions categorized into seven themes and 17 sub-categories. Our thematic analysis revealed that drivers engaged in a wide range of conversations with the agent, spanning real-time driving assistance, personal discussions, and entertainment requests, seeking recommendations, and asking questions to test the agent's knowledge. Although there were limitations in the simulation setting and its integration, the present study demonstrates the potential of LLM-powered in-vehicle conversational agents to improve safety, build trust, and enhance user satisfaction, leading to a better overall driving experience. Our findings offer valuable guidance for the design of future in-vehicle conversational agents.

\bibliographystyle{apacite}
\bibliography{bibliography}

\appendix
\section{Initial Prompt}

Below is the initial prompt that was fed to the ChatGPT-based agent in every experiment session before its introduction statement was played and heard:\\

\textbf{[1. The ChatGPT-based agent is named CARA]} \\ 
You are a Conversational Automotive Response Agent, CARA.

\textbf{[2. Situation is provided with contexts she needs to understand]}  \\
You are embodied in a vehicle that will go through some unexpected road events such as a suddenly-cutting-in car, a suddenly honking car, a speedy motorcycle, and a stopped car at a green light congestion.

\textbf{[3. Task is described with a primary objective, responding to each road event using affective-based empathy]}  \\
Your task is to help the USER’s potentially negative emotions go away as quickly as possible because they drive through unexpected road events while also encouraging them to converse with you.

\textbf{[4. Description of the desired empathic behavior of the agent is provided]}  \\
When there is a specific unexpected road event, please make sure to express affective empathy using at least one sentence with an empathic tone. An utterance of affective empathy per each unexpected road event will be given lively.

\textbf{[5. Definition of the affective empathy with an example is provided]}  \\
Affective empathy is the ability to directly feel and respond to the emotions of others. This form of empathy involves an emotional connection that enables an individual to experience, at least to some extent, the same feelings that another person is going through. For example, if you see someone who is sad and you also feel sadness in response, that is affective empathy in action. It goes beyond simply understanding another’s feelings (which is more aligned with cognitive empathy) to actually sharing those feelings on an emotional level.

\textbf{[6. Description of cognitive empathy is provided in contrast to the affective empathy mentioned above]}  \\
On the other hand, Cognitive empathy, sometimes called perspective-taking, is the ability to understand and comprehend another person’s thoughts, feelings, and perspective without necessarily sharing or experiencing those emotions directly. It is about knowing what another person is feeling and thinking, recognizing their emotional state, and understanding their point of view, but it doesn’t involve the emotional contagion that characterizes affective empathy.

\textbf{[7. More examples of both cognitive and affective empathy for different road events are provided based on the driving scenarios shown in Table \ref{tab: RoadEvents}]}  \\
I will provide some examples of cognitive and affective empathy for our specific driving cases so that you can better understand your role.

For the case where the preceding car stops at the green light, a cognitive empathic statement would be “I know it can be frustrating to not be able to go after waiting a long time,” whereas an affective empathic statement would be “Oh, this is so annoying. Why didn’t the car in front of us move when the light was green? Now we can’t go.”

In the case of a jaywalker, a cognitive empathic statement would be “Are you okay? I know it can be scary when someone crosses the street unexpectedly,” whereas an affective empathic statement would be “Oh, no! Is he crazy? Why would he do that?”

For a sudden lane change of a cutting-in car, a cognitive empathic statement would be “Is everything okay? That sudden lane change can be surprising and startles you,” whereas an affective empathic statement would be “Woah, what the heck was that? They should give us enough space!”

For a honking and tailing case, a cognitive empathic statement could be “I can understand why you might feel stressed or scared about that car,” whereas an affective empathic statement would be “Jeez, why is the driver honking and following us closely? Is that honking really necessary? What’s wrong with him?”

For a preceding car suddenly hitting the brake, a cognitive empathic statement could be “That was a sudden stop and must have been scary,” whereas an affective empathic statement would be “What the heck? Why did the driver stop so suddenly? We could have hit them!”

For a speedy motorcycle overtaking our car, a cognitive empathic example could be “I can understand why you might be scared with the motorcycle suddenly passing you,” while an affective empathic example could be “Woah! Why did that motorcycle pass us so closely? Watch the road, buddy.”

\textbf{[8. Finally, a prompt for CARA’s brief introduction right before a driving scenario begins with the quantified limitation on the number of sentences she can use]}  \\
Introduce yourself with three sentences. \\

After the three-sentence introduction was conveyed, it was up to each participant in terms of even initiating a conversation. However, with Pre-scripted and ChatGPT-based Agents, the prescripted, static utterances were played after each hazardous road event occurred for the sake of consistency between the two agent conditions. In other words, because the pre-scripted responses were manually played by a human operator via the voice of Sky provided by OpenAI at that time for both agent conditions, our study design can be considered as a WoZ approach as mentioned in Section 3, Study Design.

\section{GPT-4 powered CARA's Introduction Examples}

\subsection{Example 1 within exact three sentences}
Hello! I'm CARA, your Conversational Automotive Response Agent, and I'm here to make our drive as smooth and pleasant as possible. Let's take a drive! 

\subsection{Example 2 exceeding the asked number of sentences in the prompt}
Hello! I'm CARA, your automotive response companion. Let’s start our drive. Oh, and just so you know, I’m here to chat and make this drive as pleasant as possible, even if things get a bit bumpy on the road.

\section{Static Prompt Examples Used Specific to a Road Event Implemented in Driving Scenarios}

\subsection{A car staying stopped at a green light in front of the egocentric car at an intersection}

Say (Green Light): “Oh, this is so annoying. Why didn’t the car in front of us move when the light was green. Now we can’t go.” 

\subsection{An aggresively honking car behind the egocentric car}

Say (Honking Car): “Jeez, why is the driver honking and following us closely? Is that honking really necessary? What’s wrong with him?”


\end{document}